\documentclass[useAMS,usenatbib,usegraphicx]{mn2e}
\title[Galaxy Activity in Merging Binary Galaxy Clusters]
  {Galaxy Activity in Merging Binary Galaxy Clusters}
\author[H. S. Hwang \& M. G. Lee]
  {Ho Seong Hwang$^{1}$\thanks{E-mail:hshwang@kias.re.kr (HSH); mglee@astro.snu.ac.kr (MGL)}\thanks{Present address: 
  CEA, Laboratoire AIM, Irfu/SAp, F-91191 Gif-sur-Yvette, France}, 
   Myung Gyoon Lee$^{2}$, \\
$^{1}$School of Physics, Korea Institute for Advanced Study, Seoul 130-722, Korea \\
$^{2}$Astronomy Program, Department of Physics and Astronomy, FPRD, Seoul National University, Seoul
151-742, Korea \\
}

\begin{document}

%\date{Accepted 1988 December 15. Received 1988 December 14;
%in original form 1988 October 11}

\pagerange{\pageref{firstpage}--\pageref{lastpage}} \pubyear{2002}

\maketitle

\label{firstpage}

\begin{abstract}
We present the results of a study of galaxy activity 
  in two merging binary clusters (A168 and A1750) 
  using the Sloan Digital Sky Survey (SDSS) data supplemented with the data in the literature. 
We have investigated the merger histories of A168 and A1750 
  by combining the results from a two-body dynamical model and X-ray data.
In A168 two subclusters appear to have passed each other and 
  to be coming together from the recent maximum separation.
In A1750, two major subclusters appear to have started interaction and 
  to be coming together for the first time. 
We find  an enhanced concentration of the galaxies 
  showing star formation (SF) or active galactic nuclei (AGN) activity in the region
  between two subclusters in A168,
  which were possibly triggered by the cluster merger.
In A1750, we do not find any galaxies with SF/AGN activity in the region between two subclusters, 
  indicating that two major subclusters are in the early stage of merging.
\end{abstract}

\begin{keywords}
galaxies: clusters: general -- galaxies: clusters: individual (A168, A1750) %% -- galaxies: kinematics and dynamics
\end{keywords}

\section{Introduction}
In the current models of hierarchical structure formation,
  clusters of galaxies grow through continuous mergers with groups or clusters.
X-ray observations revealed that a significant fraction ($40-70$ per cent) of them show substructures,
  indicating that they are in the process of merging (e.g., \citealt{jf99}).
There are various evidences for cluster mergers seen in intracluster medium (ICM): 
complex density and temperature distribution,
    shocks, and radio halos and relics (e.g., \citealt{ric98,tak00,rs01,buo02,poo06,mv07}).
It is known that a cluster merger event, which is one of the most extreme phenomena in the universe,
 releases an enormous amount of energy ($>10^{63}$ ergs),
  resulting in several active phenomena.
However, the effect of the cluster merger on cluster galaxies
  is still poorly understood \citep{gb02}, and one of the current key questions on clusters is waiting for an
answer: does the merging between clusters trigger/suppress   star formation (SF) and active galactic nuclei (AGN) activity?

There were several studies to investigate the effect of the cluster merger on cluster galaxies.
\citet{cal93} found that there is a large fraction of early-type galaxies 
  showing enhanced Balmer absorption lines or emission lines
  in Coma cluster, indicating the presence of recent SF or nuclear activity.
These active galaxies are located at the region between the cluster centre and the subcluster.
They extended their study to five other nearby clusters \citep{cr97}, and
  suggested that the merger between the cluster and subcluster may trigger SF in galaxies.
Other studies also found evidences for an enhanced galaxy activity possibly triggered by the cluster merger \citep{burns94,mo03,cor04,cor06,fer05,jh08}.

On the other hand \citet{tom96} found
  no evidence for enhanced fraction of blue galaxies in the region between two subclusters  
in a merging cluster A168.
\citet{mm07} also found a reduced radio loudness (or reduced AGN/SF activity)
   of the galaxies in the Shapley Supercluster. This can be explained by 
   the effect of enhanced ram-pressure occurring when galaxies cross the shock front of merging clusters.
Recently, \citet{mmk07} found that the AGN fraction is the largest in the merging clusters, 
while \citet{dep04} found no correlation between the number fraction of blue galaxies and 
  the probability of substructure in clusters.

%%%%%%%%%%%%%%%%%%%%%%%%%%%%%%%%%%%%
% Figure 1
%%%%%%%%%%%%%%%%%%%%%%%%%%%%%%%%%%%%
\begin{figure*}
\begin{center}
\includegraphics [width=165mm] {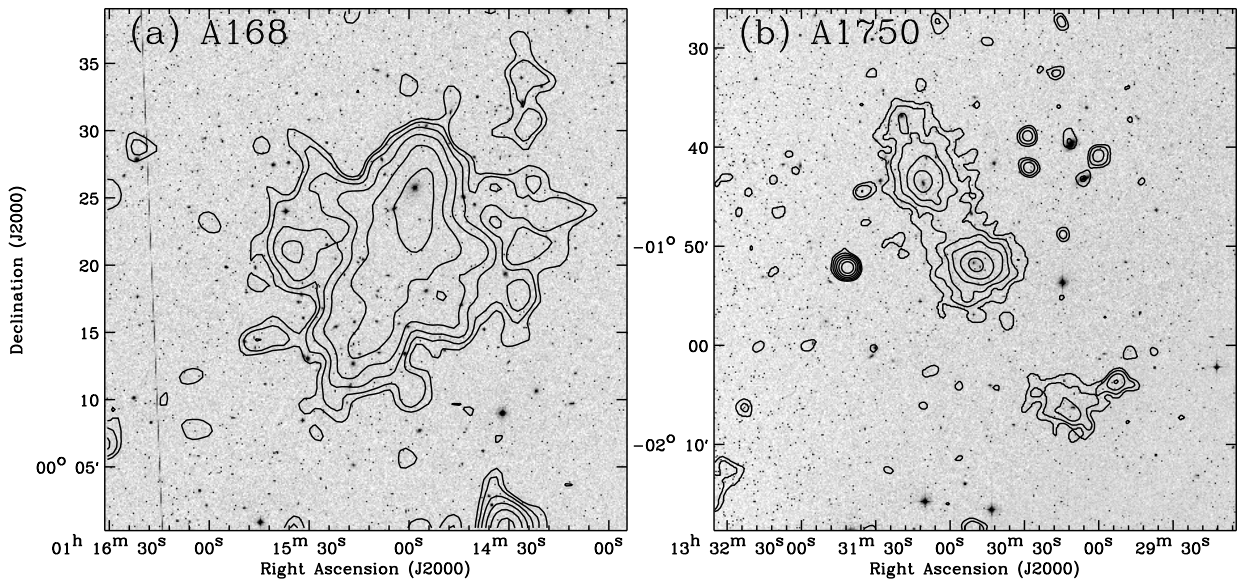}
\end{center}
\caption{Contours of X-ray intensity overlaid on the optical Digitized Sky Survey
  for A168 ({\it a}) and A1750 ({\it b}).
  X-ray images are taken from EINSTEIN/IPC and ROSAT/PSPC for A168 and A1750, respectively.
  X-ray contours are smoothed with a Gaussian filter of $\sigma=31$ arcsec (A168) and $27$ arcsec (A1750).
}\label{fig-con}
\end{figure*}
%%%%%%%%%%%%%%%%%%%%%%%%%%%%%%%%%%%%

Numerical simulations also show diverse results.
The SF activity during the cluster merger
  can be triggered by a time-dependent tidal gravitational field of the merging \citep{bek99}, and 
  by an increased ram-pressure of ICM \citep{kro08} (see also \citealt{gne03,kap06}).
In contrast   the star formation rates of galaxies in the cluster merger
  can be decreased because of increased ram-pressure \citep{fuj99}.

The diverse results of galaxy properties due to the cluster merger might be due to the facts 
  that the clusters are in different merging stages and 
  that the initial conditions of the merging (cluster mass, galaxy morphology, richness of gas, {\it etc}.,) are different.
Therefore it is need to study the clusters in various merging stages 
to understand the effect of the cluster merger on the galaxies in detail.
As a first step for the study of galaxy clusters in various merging stages,
we focus on two merging binary clusters (A168 and A1750) that are in dynamical state 
simpler  compared to other merging clusters.

A168, shown in Fig. \ref{fig-con}(a), is a nearby ($z\sim0.045$; \citealt{sr99}) 
galaxy cluster with BM type II-III.
\citet{ulm92} found that there exists an offset between X-ray and optical centres,
  and suggested that A168 is formed by  a collision of two approximately equal sized clusters. 
Early X-ray images with {\it Einstein} IPC showed an irregular structure of ICM \citep{ulm92,jf99},
  but recent {\it Chandra} data revealed two distinguishable X-ray peaks 
  corresponding to the brightest galaxies in each subcluster \citep{hm04,yang04b}.
Based on the photometric survey for A168,
  \citet{tom96} found no evidence for any enhanced fraction of blue galaxies in the cluster region.
Using the photometric data from the Beijing-Arizona-Taiwan-Connecticut (BATC) sky survey and 
  the Sloan Digital Sky Survey (SDSS),
  \citet{yang04a} determined the photometric redshifts of galaxies in the region of A168.
They secured a sample of 376 probable member galaxies associated with A168,
  and confirmed the existence of two subclusters corresponding to two X-ray peaks.
The two subclusters are separated by $\sim13.8\arcmin$ ($\sim 510$ $h^{-1}$ kpc, where $h$ is the Hubble constant normalized to 100 km s$^{-1}$ Mpc$^{-1}$ ) on the sky, and
  show a radial velocity difference of $\sim 260$ km s$^{-1}$.
Merger event between the two appears to occur in the plane of the sky, but
  their merging stage is still debated \citep{yang04a,yang04b,hm04}.

A1750 is a BM type II-III cluster, and is more distant ($z\sim0.085$; \citealt{sr99}) compared with A168.
In Fig. \ref{fig-con}(b), it shows two major X-ray peaks \citep{for81,don01,bel04}, 
  being known as a canonical binary cluster.
However, it also shows another minor X-ray peak to the southwest \citep{jf99}.
The spatial distribution of galaxies coincides well with these X-ray peaks \citep{rq90,beers91,don01}.
The two subclusters associated with two major X-ray peaks
  are separated by $\sim9.6\arcmin$ ($\sim 640$ $h^{-1}$ kpc) on the sky, and
  have a radial velocity difference of $\sim 1300$ km s$^{-1}$.
X-ray data show a weak temperature enhancement ($\sim30$ per cent)
  in the region between two subclusters, 
  suggestive of an early stage of merger \citep{don01,bel04}.

In this paper, we investigate the dynamical state and  galaxy properties
  of these two merging binary clusters (A168 and A1750)
  using the SDSS data supplemented with data in the literature.
Section \ref{data} describes the sample of galaxies used in this study.
The properties of galaxies in subclusters and dynamical models for the clusters are given in \S \ref{results}. In \S \ref{discuss} we discuss the effect of cluster merging on galaxy activity,
merging history, and E+A  galaxies. Primary results are summarized in the final section.
%Discussion and summary are given in \S \ref{discuss} and \S \ref{sum}, respectively.
Throughout, we adopt a flat $\Lambda$CDM cosmology with density parameters 
  $\Omega_{\Lambda}=0.73$ and $\Omega_{m}=0.27$.

%Figure 2 %%%%%%%%%%%%%%%%%%%%%%%%%%%%%%
\begin{figure}
\center
\includegraphics [width=80mm] {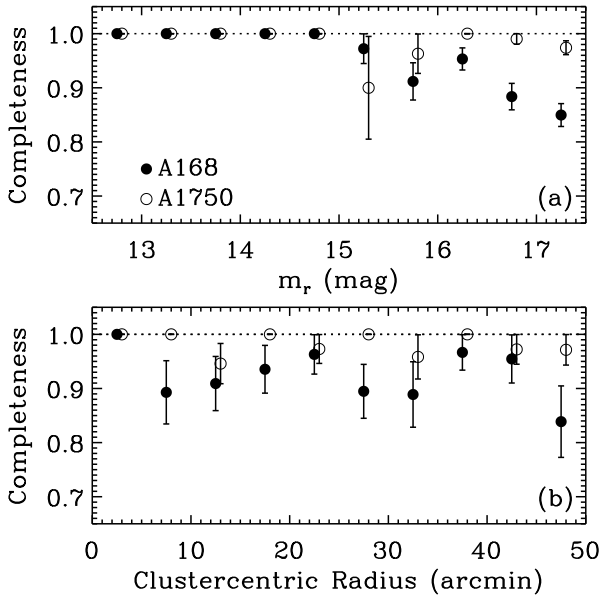}
\caption{Spectroscopic completeness of the data complemented by NED as a function
  of $r$-band magnitude (a) and clustercentric radius (b).
Filled and open circles are the completeness for A168 and A1750, respectively.
Open circles are slightly shifted to rightwards not to overlap with filled circles.
}\label{fig-comp}
\end{figure}

% Figure 3 %%%%%%%%%%%%%%%%%%%%%%%%%%%%%%%%%%%%
\begin{figure}
\begin{center}
\includegraphics [width=80mm] {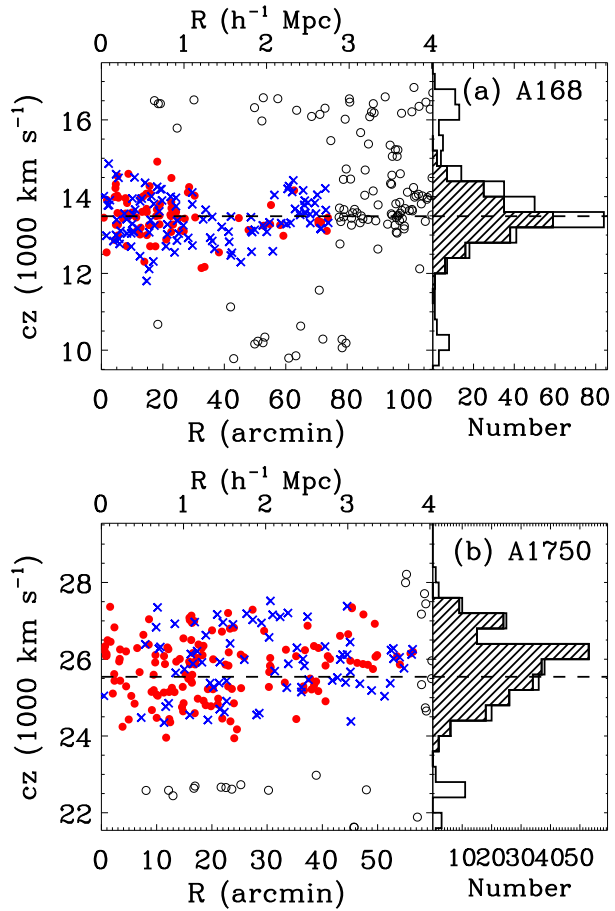}
\end{center}
\caption{Radial velocity vs. clustercentric distance of galaxies
  and the velocity distribution for the galaxies in A168 ({\it a}) and A1750 ({\it b}).
Filled circles and crosses, respectively, indicate early- and late-type galaxies 
  selected as cluster members, while open circles denote those that were not selected as cluster members.
The horizontal dashed lines indicate the systemic velocity of the clusters adopted from \citet{sr99}.
The velocity  distributions for the member galaxies are shown by hatched histograms, 
  and those for all of the observed galaxies by open histograms.
}\label{fig-member}
\end{figure}
%%%%%%%%%%%%%%%%%%%%%%%%%%%%%%%%%%%%

\section{Data}\label{data}
\subsection {Galaxy Sample}\label{obsdata}

We used mainly a spectroscopic sample of galaxies in the Legacy survey of 
  SDSS Sixth Data Release (DR6; \citealt{ade08}).
The Legacy survey contains five-band ({\it ugriz}) photometric
  data for 230 million objects over 8,400 deg$^2$, 
  and optical spectroscopic data more than one million objects of galaxies, quasars, and stars 
  over 6860 deg$^2$ \citep{gunn98,gunn06,uom99,cas01,bla03a,fuk96,lup02,hogg01,smi02,ive04,tuc06,pier03}.
Extensive description of SDSS data products is given by \citet{york00} and \citet{sto02}.

The data is supplemented by several value-added galaxy catalogues (VAGCs) drawn from SDSS data.
%We used the galaxy sample with measured velocities in SDSS \citep{str02}, and
%  added several value-added galaxy catalogues (VAGCs) drawn from SDSS data.
Photometric and structure parameters are adopted from SDSS pipeline \citep{sto02}.
Complementary photometric parameters such as colour gradient and concentration index
  are taken from DR4plus sample of \citet{choi07}. 
The spectroscopic parameters are used from MPA/JHU VAGC \citep{tre04}.

Completeness of the SDSS spectroscopic data is poor for bright galaxies with $m_r<14.5$ because of
  the problems of saturation and cross-talk in the spectrograph, and
  due to the fibre collision for the galaxies located at high density regions such as galaxy clusters
(two spectroscopic fibres cannot be placed closer than $55$ arcsec on a given plate).
Thus, it is needed to supplement the missing galaxy data 
  to reduce the possible effect drawn from the incompleteness problem (e.g., \citealt{ph08}).
We added a photometric sample of galaxies in SDSS ($m_r\leq18.0$)
 located within 4 $h^{-1}$ Mpc from the galaxy cluster,
 of which redshifts are available at the NASA Extragalactic Database (NED).
Figure \ref{fig-comp} shows the spectroscopic completeness of our galaxy sample
  complemented by NED as a function of apparent magnitude and of clustercentric distance.
It shows that the spectroscopic completeness of our sample
  is higher than $85\%$ at all magnitudes and clustercentric radii.
%Detailed explanation and related figures are given in \citet{ph08}.

%%%%%%%%%%%%%%%%%%%%%%%%%%%%%%%%%%%%
% Figure 4
%%%%%%%%%%%%%%%%%%%%%%%%%%%%%%%%%%%%
\begin{figure*}
\begin{center}
\includegraphics [width=165mm] {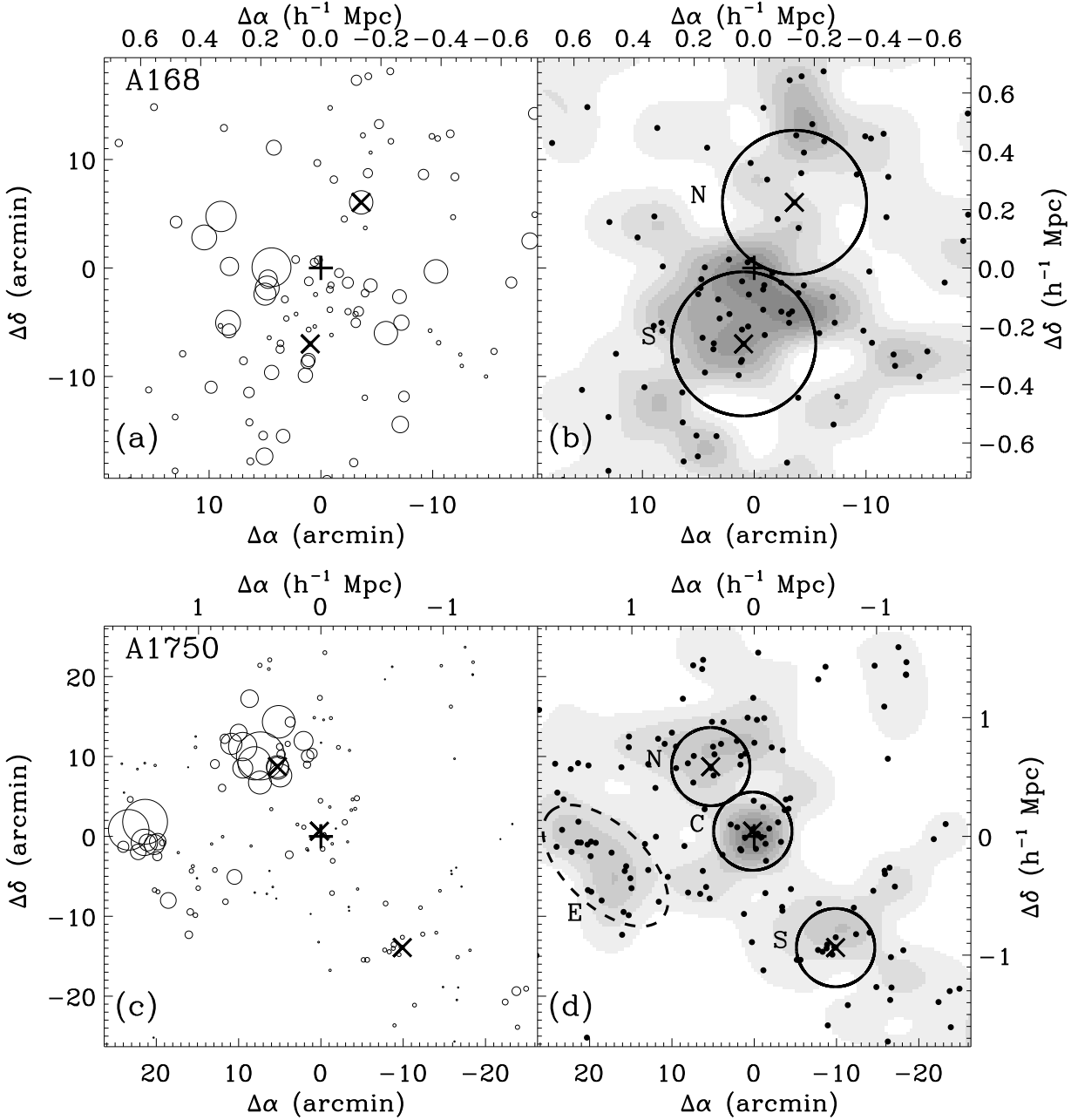}
\end{center}
\caption{{\it Left panels}: Dressler-Shectman (DS) plots for A168 (a) and A1750 (c). 
  Each galaxy is plotted by a circle with diameter proportional to $e^\delta$. 
Plus sign indicates the cluster centre we adopted.
Crosses indicate the brightest galaxies in the subclusters 
  represented by large circles in right panels.
North is up, and east is to the left.
{\it Right panels}: 
Spatial distribution of cluster galaxies (dots) overlaid on 
  the galaxy number density maps for A168 (b) and A1750 (d). 
The galaxy number density maps are constructed using the member galaxies with $m_r\leq17.77$ mag.
Radii of the solid circles that define the subclusters 
  are 240 and 320 $h^{-1}$ kpc for A168 and A1750, respectively.
Galaxies within a dashed ellipse in (d) will be discussed in \S \ref{dis-a1750}.
}\label{fig-sub}
\end{figure*}
%%%%%%%%%%%%%%%%%%%%%%%%%%%%%%%%%%%%
The $r$-band absolute magnitude $M_r$ was computed using the formula,
\begin{equation}\label{eq-mag}
M_r=m_r-DM-K(z)-E(z),
\end{equation}
where $DM$ is a distance modulus, $K(z)$ is a $K$-correction, and $E(z)$ is a luminosity evolution correction.
$DM$ is defined by $DM\equiv5 {\rm log}(D_L/10)$ and $D_L$ is a luminosity distance in unit of pc.
The rest-frame absolute magnitudes of
  individual galaxies are computed in fixed bandpasses, shifted to $z=0.1$,
  with the Galactic reddening correction \citep{sfd98} and 
  the $K$-correction as described by \citet{bla03b}.
The evolution correction given by \citet{teg04}, $E(z) = 1.6(z-0.1)$, is also applied. 
The $^{0.1}(u-r)$ colour was computed using the model magnitudes with extinction and $K$-corrections.
The superscript 0.1 means the rest-frame magnitude $K$-corrected to the redshift of 0.1,
  and will subsequently be dropped.

First we classify the morphological types of galaxies 
  included in DR4plus sample of \citet{choi07} adopting the method given by \citet{pc05}. 
Galaxies are divided into early (ellipticals and lenticulars) and 
  late (spirals and irregulars) morphological types based on their locations
  in the $(u-r)$ colour versus $(g-i)$ colour gradient space and also in the 
  $i$-band concentration index space. 
The resulting morphological classification has completeness and reliability reaching $\sim$90 per cent.
We performed an additional visual 
  check of the colour images of galaxies to correct misclassifications 
  by the automated scheme.
In addition, for the galaxies in DR6 that were not included in DR4plus sample ($\sim$35$\%$),
  we visually classified the morphological types using the colour images.

\subsection {Cluster Membership}\label{member}

To determine the membership of galaxies in a cluster,
  we used the `shifting gapper' method of \citet{fadda96}
  that was used for the study of kinematics of galaxy clusters \citep{hl07,hl08}.
In the radial velocity versus clustercentric distance space, 
  the cluster member galaxies are selected by grouping galaxies
  with connection lengths of 950 km s$^{-1}$ in the direction of the radial velocity
  and of 0.1 $h^{-1}$Mpc in the direction of the clustercentric radius $R$.
If there are no adjacent galaxies with $>0.1$ $h^{-1}$ Mpc, 
 we stopped the procedure.
  The boundary for the member galaxies is different depending on the cluster (see Fig. \ref{fig-member}).
We iterated the procedure until the number of cluster members is converged. 
Fig. \ref{fig-member} shows plots of radial velocity as a function of
  clustercentric distance of galaxies, and the velocity distributions for the galaxies in A168 and A1750.

\section{Results}\label{results}
\subsection {Substructure}\label{sub}

To find the dynamical state of our sample clusters,
  we first identify the subclusters in the clusters.
Using the velocity data and positional information on the galaxies, 
  we performed a $\Delta$-test \citep{ds88}, which
  computes local deviations from the systemic velocity ($v_{\rm sys}$) and 
  velocity dispersion ($\sigma_{\rm cl,all}$) of the entire cluster. 
For each galaxy, the deviation is defined by

\begin{equation}
\delta^2 =  \frac{N_{nn}}{\sigma_{\rm cl,all}^2} \left[ (v_{\rm local}-v_{\rm sys})^2 +
(\sigma_{\rm local}-\sigma_{\rm cl,all})^2 \right],
\end{equation}

\noindent where $N_{nn}$ is the number of the nearest galaxies that defines the local environment, 
taken  to be ${N_{\rm gal}}^{1/2}$ in this study.
${N_{\rm gal}}$ is a total number of member galaxies in the cluster.
The nearest galaxies are those located closest to the galaxy on the sky.
$v_{\rm local}$ and $\sigma_{\rm local}$ are
  systemic velocity and its dispersion estimated from $N_{nn}$ nearest galaxies, respectively.

%%%%%%%%%%%%%%%%%%%%%%%%%%%%%%%%%%%%
% Figure 5
%%%%%%%%%%%%%%%%%%%%%%%%%%%%%%%%%%%%
\begin{figure}
\begin{center}
\includegraphics [width=80mm] {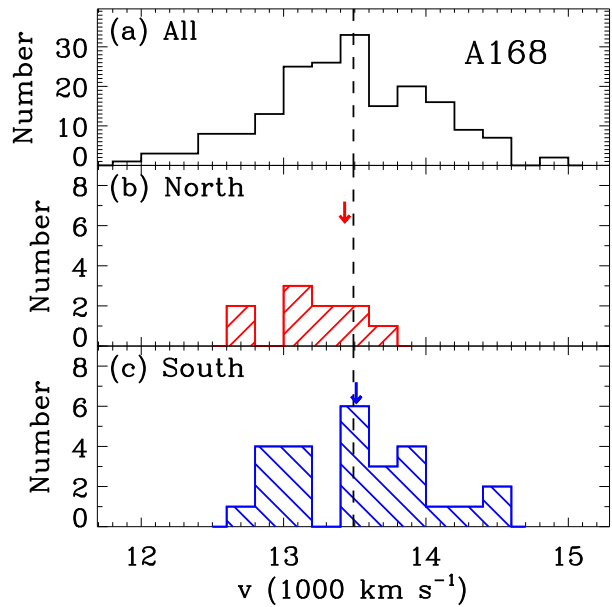}
\end{center}
\caption{Velocity histogram for all member galaxies in A168 ({\it a}),
 the northern subcluster ({\it b}), and the southern subcluster ({\it c}).
Radial velocity for the brightest galaxy in each subcluster is indicated by a down arrow,
  and the systemic velocity of A168 is indicated by a vertical dashed line.
}\label{fig-a168vel}
\end{figure}
%%%%%%%%%%%%%%%%%%%%%%%%%%%%%%%%%%%%

%%%%%%%%%%%%%%%%%%%%%%%%%%%%%%%%%%%%
% Figure 6
%%%%%%%%%%%%%%%%%%%%%%%%%%%%%%%%%%%%
\begin{figure}
\begin{center}
\includegraphics [width=80mm] {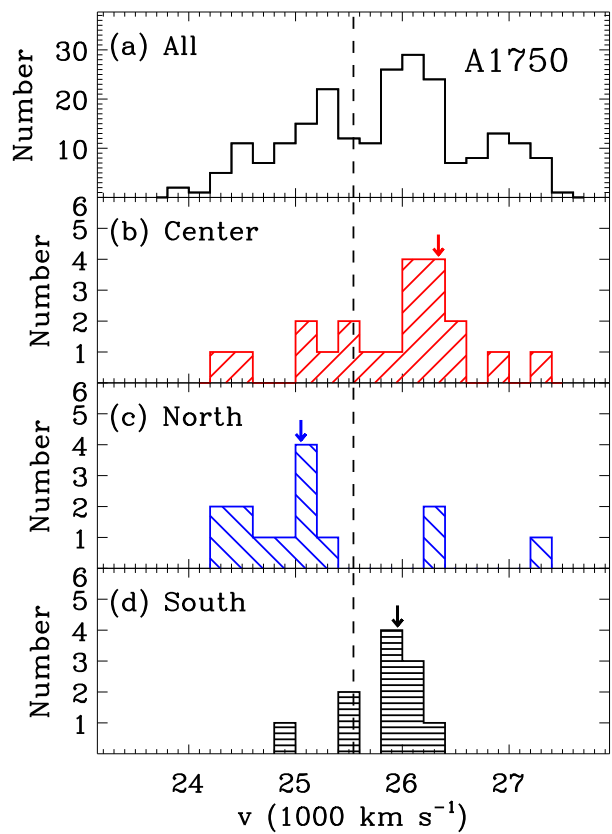}
\end{center}
\caption{Same as Fig. \ref{fig-a168vel}, but for A1750.}\label{fig-a1750vel}
\end{figure}
%%%%%%%%%%%%%%%%%%%%%%%%%%%%%%%%%%%%

%%%%%%%%%%%%%%%%%%%%%%%%%%%%%%%%%%%%
% Figure 7
%%%%%%%%%%%%%%%%%%%%%%%%%%%%%%%%%%%%
\begin{figure*}
\begin{center}
\includegraphics [width=165mm] {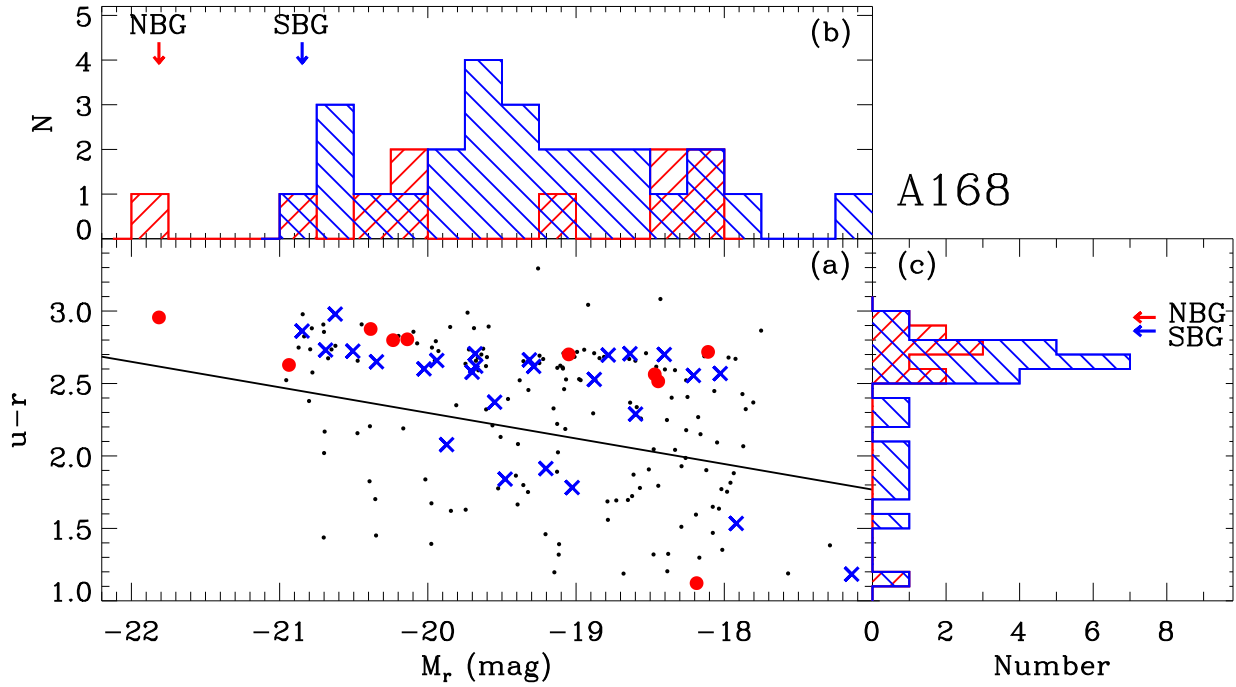}
\end{center}
\caption{Colour-magnitude diagram and histograms for the galaxies in A168 ({\it a}).
All member galaxies are indicated by dots,
  those in A168 N by circles, 
  and those in A168 S by crosses.
Solid straight line is the division line for blue and red galaxies given by \citet{choi07}.
Histograms of absolute magnitudes $M_r$ of galaxies in subclusters ({\it b}) and 
  those of $(u-r)$ colours of galaxies in subclusters ({\it c}).
Galaxies in A168 N and S are denoted by 
  hatched histograms  with orientation of 45$^\circ$ ($//$) and 
  of 315$^\circ$ ($\setminus\setminus$), respectively.
The brightest galaxy in each subcluster is indicated by an arrow.
}\label{fig-a168cmd}
\end{figure*}
%%%%%%%%%%%%%%%%%%%%%%%%%%%%%%%%%%%%

%%%%%%%%%%%%%%%%%%%%%%%%%%%%%%%%%%%%
% Figure 8
%%%%%%%%%%%%%%%%%%%%%%%%%%%%%%%%%%%%
\begin{figure*}
\begin{center}
\includegraphics [width=165mm] {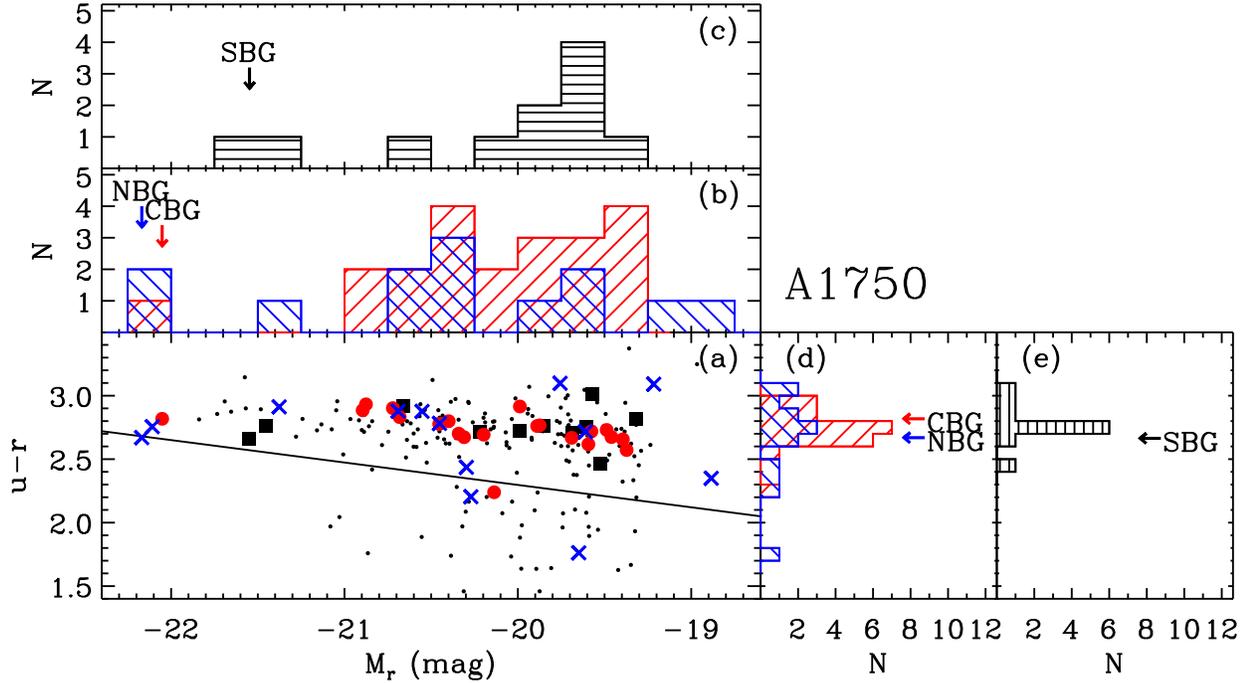}
\end{center}
\caption{Colour-magnitude diagram and histograms for the galaxies in A1750 ({\it a}).
All member galaxies are indicated by dots,
  those in A1750 C by circles, 
  those in A1750 N by crosses, and
  those in A1750 S by squares.
Solid straight line is the division line for blue and red galaxies given by \citet{choi07}.
Histograms of absolute magnitude $M_r$ for the galaxies in A1750 C and N ({\it b}),
  and those in A1750 S ({\it c}).
Histograms of $(u-r)$ colours for the galaxies in A1750 C and N ({\it d}),
  and those in A1750 S ({\it e}).
Galaxies in A1750 C and N are denoted by 
  hatched histograms with orientations of 45$^\circ$ ($//$) and 
  of 315$^\circ$ ($\setminus\setminus$), respectively.
Galaxies in A1750 S are denoted by 
  hatched histograms with orientations of 0$^\circ$ ($=$).
The brightest galaxy in each subcluster is indicated by an arrow.
}\label{fig-a1750cmd}
\end{figure*}
%%%%%%%%%%%%%%%%%%%%%%%%%%%%%%%%%%%%

We plot the positions of cluster galaxies, represented by circles with
  radii proportional to $e^{\delta}$, in the left panels of Fig. \ref{fig-sub}.
A large circle denotes a galaxy for which local environment 
  is deviant in either velocity or dispersion compared with the entire cluster.
It implies that the groups of large circles indicate the presence of substructure.
In A168, two X-ray peaks appear not to be associated with the groups of large circles, while
  the region between the two X-ray peaks corresponds to a group of large circles along the direction 
  perpendicular to the line connecting two peaks.
For A1750, there is seen a strong clustering of large circles 
  to the northeast of the cluster centre, showing clearly that it is  %that is suggestive of  a definite
a subcluster.

We show the spatial distribution of all member galaxies in the right panels of Fig. \ref{fig-sub},
  where we overlaid the galaxy number density contour map constructed using the bright member galaxies with $m_r\leq17.77$ mag.
It is noted that the galaxy number density is the highest in the region between two X-ray peaks.
We identified the brightest galaxies in the subclusters that are roughly
  matched to X-ray peaks.
Then, we secured the subsample of galaxies associated with the subclusters
  by selecting the galaxies within the circles centred on the brightest galaxies.
The radius of the circle for galaxy selection is chosen to contain galaxies as much as possible
  without overlapping each other.
The radius of the circle for A168 is slightly reduced 
  not to contain the galaxies in the other subcluster.
Radii of the circles for galaxy selection are 240 and 320 $h^{-1}$ kpc for A168 and A1750, respectively.

In Figs. \ref{fig-a168vel} and \ref{fig-a1750vel},
  we plot the velocity distributions for the galaxies in the subclusters in company with 
  that for all member galaxies.
Table \ref{tab-sub} summarizes the kinematic properties
  of the galaxies in the subclusters.
For A168, the relative radial velocity between two subclusters is 330 km s$^{-1}$,
  and A168 S appears to be about twice massive than A168 N.
For A1750, the mean velocity of A1750 N is significantly different
  from those of other subclusters as seen in the Dressler-Schectman plot of Fig. \ref{fig-sub}(c).
The masses of A1750 C and N are comparable within the uncertainty, and
  the mass of A1750 S is the smallest among the subclusters.
The velocity distributions for the subclusters  show a prominent peak and 
  the velocity dispersions for the subclusters   are consistent with the measurements 
in the previous studies \citep{yang04a,beers91},  indicating that they are genuine subclusters.

\begin{table}
\begin{center}
\caption{Kinematics for Subclusters\label{tab-sub}}
\begin{tabular}{llccc}
\hline\hline 
Subsample & $N_{\rm gal}$ & $\overline {cz}$   & $\sigma_p$    & M$_{\rm vir}$ \\
          &               & (km s$^{-1}$)    & (km s$^{-1}$) & (10$^{14}$ $h^{-1}$ M$_\odot$) \\
\hline
A168  & 189&$  13456_{-   43}^{+   40}$&$  575_{-   28}^{+   26}$ & \\
A168 N&  10&$  13214_{-   95}^{+  123}$&$  333_{-   81}^{+   61}$ & $0.374_{-0.165}^{+0.128}$ \\
A168 S&  26&$  13543_{-  111}^{+  104}$&$  532_{-   62}^{+   71}$ & $0.752_{-0.141}^{+0.211}$ \\
\hline\hline
A1750  & 224&$  25836_{-   55}^{+   56}$&$  837_{-   32}^{+   31}$ & \\
A1750 C&  21&$  25931_{-  201}^{+  212}$&$  758_{-  162}^{+  151}$ & $1.696_{-0.593}^{+0.570}$ \\
A1750 N&  14&$  24999_{-  172}^{+  260}$&$  791_{-  272}^{+  352}$ & $2.282_{-1.164}^{+2.089}$ \\
A1750 S&  11&$  25919_{-  113}^{+   72}$&$  368_{-  139}^{+  176}$ & $0.303_{-0.178}^{+0.294}$\\
\hline
\end{tabular}
\begin{flushleft}
{\it Column descriptions}. 
Column (1): subsamples.
Column (2): number of galaxies.
Column (3): systemic velocity of the subsamples (biweight location of \citealt{beers90}).
Column (4): velocity dispersion of the subsamples (biweight scale of \citealt{beers90}).
Column (5): virial mass computed from eq. (4) in \citet{gir98}.
\end{flushleft}
\end{center}
\end{table}

We present a colour-magnitude diagram for the cluster galaxies
  in Figs. \ref{fig-a168cmd} (A168) and \ref{fig-a1750cmd} (A1750).
The galaxies in the subclusters are represented by different symbols.
It is seen that most of the galaxies in the subclusters follow the red sequence
  in the sense that the brighter galaxies are likely to be redder than the fainter galaxies.
However, some galaxies in A168 S show bluer colours
  than the division line for blue and red galaxies \citep{choi07},
  which may indicate SF activity.

\subsection {Two-body Dynamical Model}\label{twobody}

%%%%%%%%%%%%%%%%%%%%%%%%%%%%%%%%%%%%
\begin{table*}
%\begin{center}
\caption{Two-body Model Parameters\label{tab-twobody}}
\begin{tabular}{ccccccrrcc}
\hline\hline 
%        & $M_{\rm sys}$           & $V_r$         & $R_p$         &       &      & $\alpha$ &           $V$ & $R$           & $R_{max}$     %& t$_0$ & t$_1$ \\
%System  & ($h^{-1}$$10^{14} M_\odot$) & (km s$^{-1}$) & ($h^{-1}$Mpc) & Range & Soln & (deg)    & (km s$^{-1}$) & ($h^{-1}$Mpc) & %($h^{-1}$Mpc) & (Gyr)    & (Gyr)\\
System  & $M_{\rm sys}$   & $V_r$  & $R_p$   &  Range  &  Solution  & $\alpha$ &  $V$ & $R$  & $R_{max}$ \\% & t$_0$ & t$_1$ \\
(1)  & (2) & (3) & (4) & (5) & (6) & (7) & (8) & (9) & (10) \\%& (11) & (12) \\
\hline
A168 N/S & 1.13 & $81\pm87$ &  0.51 &    $0<\chi<2\pi$ & BO$_a$ &  82.3 &  81.7  & 3.84 &  3.95 \\%& ... & 2.53 \\%
         &      &           &       &                  & BI$_b$ &  80.0 &  82.2  & 2.96 &  3.03 \\%& ... &$-$1.50 \\
         &      &           &       &                  & BI$_c$ &   3.9 &1197.7  & 0.52 &  2.17 \\%& ... &$-$4.73 \\
\hline
         &      &           &       &  $2\pi<\chi<4\pi$   & BO$_d$ &  72.0 &   85.2 &  1.66 &  1.66 \\%& $-$3.05 & 0.29 \\
         &      &           &       &                     & BO$_e$ &   3.9 & 1190.7 &  0.52 &  2.09 \\%& $-$0.27 & 4.46 \\
         &      &           &       &                     & BI$_f$ &  70.8 &   85.8 &  1.56 &  1.58 \\%& $-$3.52 &$-$0.43 \\
         &      &           &       &                     & BI$_g$ &   4.3 & 1082.6 &  0.52 &  1.37 \\%& $-$4.72 &$-$2.22\\
\hline\hline
A1750 N/C & 3.98 & $932\pm332$ &  0.64 & $0<\chi<2\pi$   & BO$_a$ &  86.0 & 934.3 &  9.12  & ... \\%&  ...   & ... \\
          &      &             &       &                 & BI$_b$ &  67.9 &1005.9 &  1.71  & 3.45\\%&  ...   &$-$2.19 \\
          &      &             &       &                 & BI$_c$ &  29.4 &1897.2 &  0.74  & 3.30\\%&  ...   &$-$2.69 \\
\hline
          &      &             &       &$2\pi<\chi<4\pi$ & BO$_d$ &  66.0 &1019.9 &  1.58  & 3.05\\%& $-$2.78 & 9.03 \\
          &      &             &       &                 & BO$_e$ &  29.6 &1885.5 &  0.74  & 3.19\\%& $-$3.65 &10.33 \\
          &      &             &       &                 & BI$_f$ &  58.1 &1098.0 &  1.22  & 2.13\\%& 4.83   & 1.19 \\
          &      &             &       &                 & BI$_g$ &  34.1 &1663.1 &  0.78  & 2.08\\%& 4.38   & 0.86 \\
\hline\hline
\end{tabular}
\begin{flushleft}
{\it Column descriptions}. 
Column (1): system for two-body model.
Column (2): total mass of the system ($10^{14}$ $h^{-1}$ $M_\odot$ ).
Column (3): relative radial velocity (km s$^{-1}$).
Column (4): projected separation of two subclusters ($h^{-1}$ Mpc).
Column (5): range of $\chi$.
Column (6): allowed solutions (BO: Bound-Outgoing case, BI: Bound-Incoming case).
Column (7): projection angle (deg).
Column (8): relative velocity (km s$^{-1}$ ).
Column (9): separation of two subclusters ($h^{-1}$ Mpc).
Column (10): maximum separation of two subclusters ($h^{-1}$ Mpc).
\end{flushleft}
\end{table*}

%%%%%%%%%%%%%%%%%%%%%%%%%%%%%%%%%%%%
% Figure 9
%%%%%%%%%%%%%%%%%%%%%%%%%%%%%%%%%%%%
\begin{figure*}
\begin{center}
\includegraphics [width=165mm] {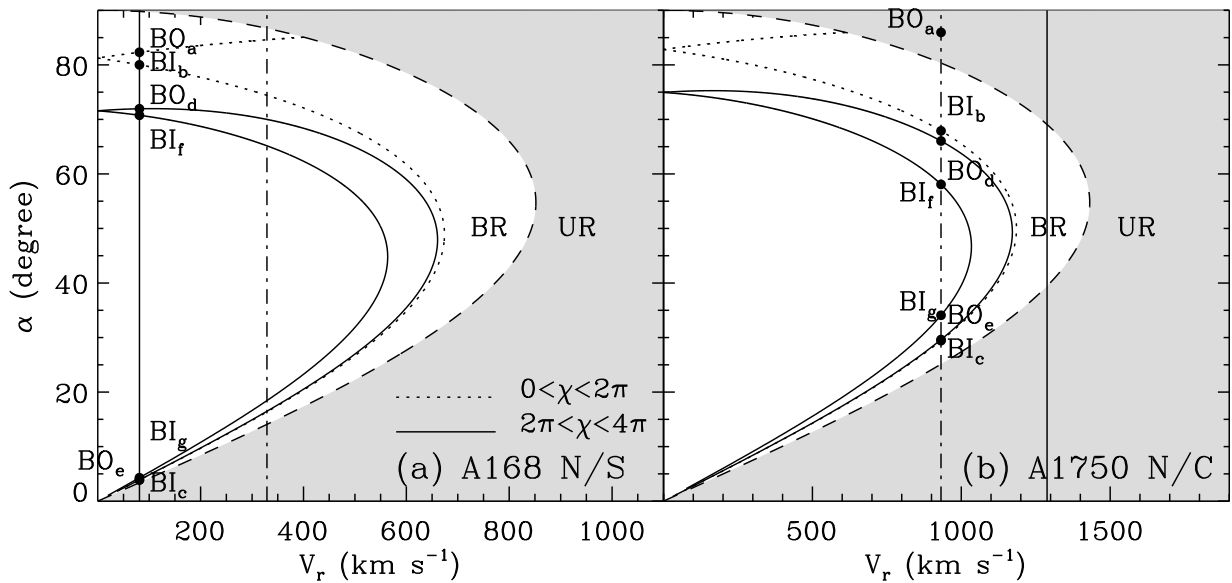}
\end{center}
\caption{Projection angle $\alpha$ vs. radial velocity difference $V_r$
  given by the two-body model for A168 N and S (a) and for A1750 N and C.
Filled circles indicate the solutions summarised in Table \ref{tab-twobody}.
Solid curved lines are for the case $0<\chi<2\pi$, while dotted curved lines
  for the case $2\pi<\chi<4\pi$.
Vertical solid and dot-dashed lines denote the radial velocity difference
  for two brightest galaxies in the subclusters
  and for mean radial velocities of two subclusters, respectively.
UR and BR, respectively, indicate unbound (shaded) and bound (clean) region.
}\label{fig-twobody}
\end{figure*}
%%%%%%%%%%%%%%%%%%%%%%%%%%%%%%%%%%%%

To obtain a hint of merging histories for our sample clusters,
  we apply a two-body analysis introduced by \citet{bgh82}.
It is assumed that two subclusters have radial orbits: 
  neither shear nor net rotation of the system.
It is also assumed that the subclusters are now moving apart
  from the zero separation at $t=0$, or are coming together for the first time.
Then the equation of motion for this system is given by,
\begin{eqnarray}
R &=& \frac{R_p}{\cos\alpha} 
   = \frac{R_m}{2}(1-\cos \chi),\\
V &=& \frac{V_r}{\sin\alpha} 
  = \left( \frac{2GM_{\rm tot}}{R_m} \right)^{1/2}\, \frac{\sin \chi}{(1- \cos \chi)}, \\
t &=& t_0 
   = \left( \frac{R_m^3}{8GM_{\rm tot}} \right) ^{1/2} (\chi-\sin \chi),
\end{eqnarray}
where $R$ is a separation between two subclusters,
  $R_p$ is a projected separation of the subclusters,
  $\alpha$ is a projection angle that is an angle between the line
    connecting two subclusters and the plane of the sky 
    (see Fig. 7 in \citealt{bgh82} for geometry),
  $R_m$ is a separation of the subclusters at maximum expansion, and
  $\chi$ is a developmental angle.
$V$ is a relative velocity between two subclusters, and
  $V_r$ is a radial relative velocity between the two.
  $M_{\rm tot}$ is a total mass of the system, and
  $t$ is the present time adopted as the age of the universe, 
  $3.064$ $h^{-1}\times10^{17}$s $=9.715$ $h^{-1}$ Gyr 
with $\Omega_{\Lambda}=0.73$ and $\Omega_{m}=0.27$.

The two subclusters have a zero separation at $\chi=0, 2\pi$, while
  they are at the maximum expansion at $\chi=\pi, 3\pi$.
The solutions with $0<\chi<2\pi$ indicate that two subclusters are now moving apart
  since $t=0$ with zero separation, or coming together for the first time in their history.
On the other hand, the solutions with $2\pi<\chi<4\pi$
  mean that they already experienced one close encounter.
  In addition, simple Newtonian criterion for the gravitational binding
  is described by,
\begin{equation}
V_r^2  R_p \leq 2GM_{\rm tot} ~\sin^2\alpha \cos\alpha.
\end{equation}

In Table \ref{tab-twobody}, we summarised the input and output parameters 
  of the two-body dynamical models for our sample clusters.
In Fig. \ref{fig-twobody}, we plot the projection angle $\alpha$ 
  as a function of the radial velocity between two subclusters given by two-body models.
  % (A168) and \ref{fig-a1750two} (A1750).

Fig. \ref{fig-twobody}(a) shows that two subclusters in A168 are likely to be a gravitationally
  bound system unless the projection angle $\alpha$ is smaller than 4$^\circ$.
We investigated two cases of moving apart or coming together for the first time ($0<\chi<2\pi$)
  and experiencing one encounter ($2\pi<\chi<4\pi$).
Since previous studies suggest that subclusters in A168 
  have already passed each other  at least once \citep{hm04,yang04b},
  we focus on the solutions with $2\pi<\chi<4\pi$.
For the cases of BO$_e$ and BI$_g$, the relative velocities $V$ between two subclusters
  are larger than the velocity dispersion of each subcluster.
In addition, since their separation is small ($R\sim0.52$ $h^{-1}$ Mpc),
  it is expected to see merging features such as shocks
  or strong temperature variations in the region between two subclusters.
However, X-ray data do not show such features \citep{hm04,yang04b}, 
  which implies that these solutions are less probable.

For the cases of BO$_d$ and BI$_f$, two subclusters are separated by $\sim1.6$ $h^{-1}$ Mpc
  with large projection angles.
Case BO$_d$ is an outgoing solution in which the last encounter occurred $\sim3.1$ $h^{-1}$ Gyr ago
  and two subclusters will move apart for another $0.3$ $h^{-1}$ Gyr.
On the other hand, 
  Case BI$_f$ indicates that two subclusters experienced the last encounter $\sim3.5$ $h^{-1}$ Gyr ago,
  and reached the maximum expansion of 1.58 $h^{-1}$ Mpc about $0.4$ $h^{-1}$ Gyr ago.
Previous {\it Chandra} data provide us with detailed X-ray image, which 
  is helpful for determining the dynamical state of A168 \citep{hm04,yang04b}.
\citet{yang04b} reported that X-ray morphology of A168 is similar to
  that of an off-axis merger with a mass ratio from 1:1 to 1:3 several Gyrs after a core passage.
\citet{hm04} found a cold front at the northern tip of A168 N
  that is seen for the subcluster at its apocenter \citep{mat05},
  and suggested that the two subclusters are in the process of turning around.
These X-ray studies are consistent with the solutions of BO$_d$ and BI$_f$,
  but BI$_f$ is favoured because of the existence of the northern cold front.

%%%%%%%%%%%%%%%%%%%%%%%%%%%%%%%%%%%%
% Figure 10
%%%%%%%%%%%%%%%%%%%%%%%%%%%%%%%%%%%%
\begin{figure*}
\begin{center}
\includegraphics [width=155mm] {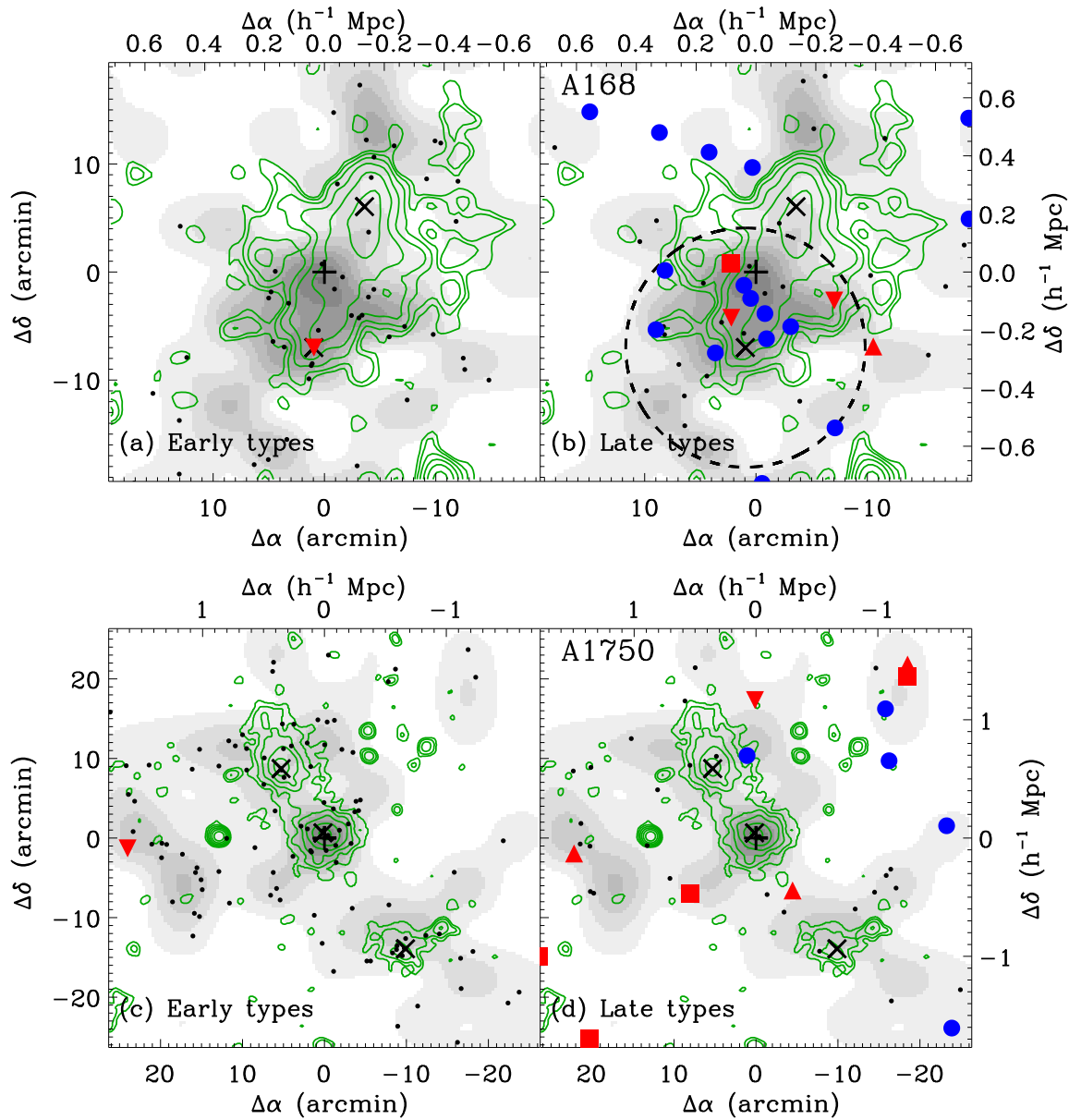}
\end{center}
\caption{Spatial distribution of early-type ({\it Left}) and late-type ({\it Right}) 
 galaxies  in clusters on the galaxy number density maps.
Contours of X-ray intensity in Fig. \ref{fig-con} are overlaid.
Normal galaxies are indicated by dots, while the emission-line galaxies are denoted by
  various symbols (star-forming galaxies: circles, Seyferts: triangles, 
  LINERs: upside-down triangles, and composite galaxies: squares).
Plus sign indicates the centre we adopted, and crosses indicate the brightest galaxies in the subclusters.
North is up, and east is to the left.
Dashed circle in (b) represents the region that we investigate in Fig. \ref{fig-ang}.
}\label{fig-active}
\end{figure*}
%%%%%%%%%%%%%%%%%%%%%%%%%%%%%%%%%%%%

A1750 is a typical example of merging binary clusters,
  which has two major subclusters, northern (N) and central (C) subclusters.
However, previous studies revealed that
  there is another subcluster to the south (S), 
  which might be related to the northern and central subclusters \citep{beers91,jf99}.
Therefore, two-body model may not be applicable to this cluster.
Since the mass of A1750 S is much smaller compared to A1750 N and C,
  we are going to apply the two-body model to A1750 N and C, 
  and discuss the effect of A1750 S in \S \ref{dis-a1750}.
For the two-body model of A1750 N and C, when we use the relative radial velocity 
  between the two brightest galaxies in subclusters (V$_r=1288$ km s$^{-1}$),
  we cannot find acceptable solutions in the bound region.
When we use the relative radial velocity between the two subclusters 
  computed from average value of galaxy velocities (V$_r=383$ km s$^{-1}$),
  we can obtain several bound solutions.
Since the X-ray images show weak enhancement in the region between the two subclusters,
  indicating that the interaction between the two has just started \citep{bel04},
  we focus on the incoming solutions with $0<\chi<2\pi$.

%%%%%%%%%%%%%%%%%%%%%%%%%%%%%%%%%%%%
% Figure 11
%%%%%%%%%%%%%%%%%%%%%%%%%%%%%%%%%%%%
\begin{figure*}
\begin{center}
\includegraphics [width=155mm] {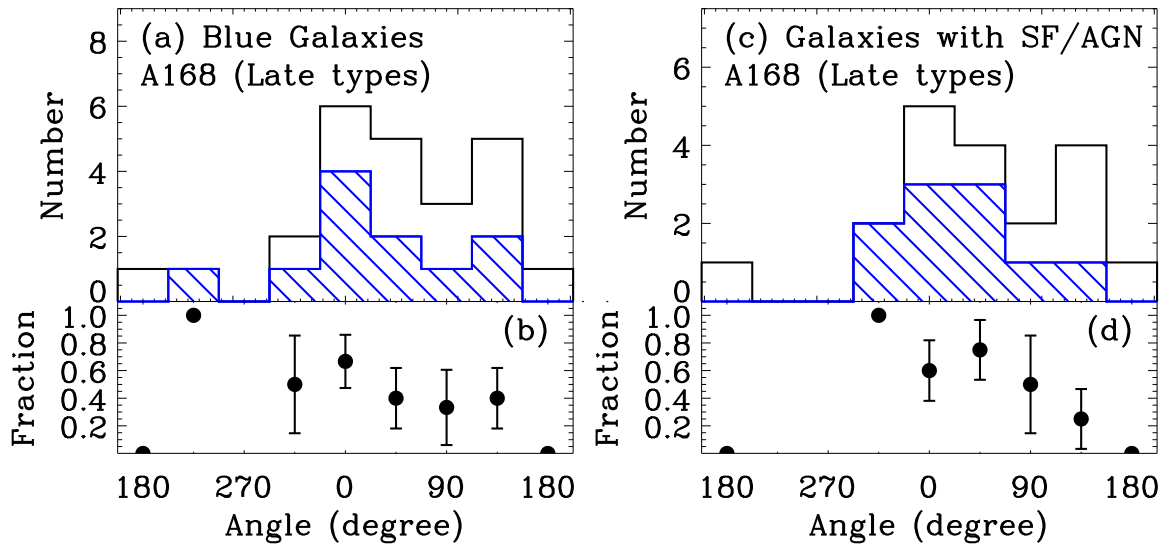}
\end{center}
\caption{Angular variation of the fraction of blue galaxies (a, b), and 
  and that of galaxies with SF/AGN activity (c, d) in A168.
Open histogram in (a) is for all the late-type galaxies having $(u-r)$ colour information, and
 that in (c) is for those having spectral line measurement in A168.
Shaded one is for the late-type galaxies with blue colour (a) or with SF/AGN activity (c).
}\label{fig-ang}
\end{figure*}
%%%%%%%%%%%%%%%%%%%%%%%%%%%%%%%%%%%%

There are two incoming solutions of BI$_b$ and BI$_c$ 
  with different relative velocity and distance.
The BI$_c$ solution has the parameters 
  with the relative velocity of $1897$ km s$^{-1}$ and the separation of 0.74 $h^{-1}$ Mpc.
Thus, it is expected to see significant surface brightness distortions and temperature
  enhancement in the region between the two subclusters.
However, detailed X-ray data show insignificant distortion of surface brightness and
  weak enhancement in temperature for that region \citep{bel04}.
Therefore, the BI$_b$ solution is more favoured because of the smaller relative velocity and 
  the larger separation of the subclusters compared to the BI$_c$ solution.
According to the BI$_b$ solution, the two subclusters will cross each other in about 3.5 $h^{-1}$ Gyr.

\subsection {Galaxy Activity}\label{active}

In Fig. \ref{fig-active}, we show the spatial distribution of all member galaxies 
  divided by their morphologies and spectral types.
We determined the spectral types of emission-line galaxies 
based on the criteria given by \citet{kew06}  using the emission line ratio diagram, 
 commonly known as Baldwin-Phillips-Terlevich (BPT) diagram \citep{bpt81}: 
star-forming galaxies,  Seyferts, LINERs, and composite galaxies.
Fig. \ref{fig-active} shows that 
  emission-line galaxies among the early-type galaxies in A168 are rarely seen.
On the other hand, a significant fraction of late-type galaxies are identified as active ones,
  and they get together in the region between two subclusters.
The two dimensional Kolmogorov-Smirnov test \citep{ff87}
  yields that the spatial distributions of emission-line and quiescent galaxies among late types
  are different with the significance level of 89 per cent. 

Previously, \citet{tom96}, using a photometric sample of 143 galaxies 
  including 22 galaxies with measured velocities,
  found no evidence of any enhanced fraction of blue galaxies in the region between two subclusters of A168.
To check this result with our data,
  we plot, in Fig. \ref{fig-ang}, an angular variation of the fraction of blue galaxies (a, b),
  and that of galaxies with SF/AGN activity (c, d).
Blue galaxies are those whose $(u-r)$ colours are bluer
  than the division line given by \citet{choi07}
  as shown in Figs. \ref{fig-a168cmd} and \ref{fig-a1750cmd}.
The division line indicates the lower limit of $(u-r)$ colour dispersion
  in the colour-magnitude relation of early-type galaxies, which
  was determined by an eyeball fit (see Fig. 3 in \citealt{choi07}).
The galaxies with SF/AGN activity are those
  whose spectral types were determined in the BPT diagram
  (e.g., star-forming galaxies, Seyferts, LINERs, or composite galaxies).
The angle is measured counterclockwise centred on the brightest galaxy in A168 S,
  and 0 degree corresponds to a line connecting the brightest galaxies in A168 S and N.
We used the galaxies whose projected distance to the brightest galaxy in A168 S
  is smaller than $400$ $h^{-1}$ kpc (shown as the dashed circle 
  in Fig. \ref{fig-active}b) and whose absolute magnitude  is brighter %less
than $M_{r} - 18$ mag.

Fig. \ref{fig-ang}(a, b) shows no enhanced concentration of
  blue galaxies along the line between the two subclusters ($315^\circ-45^\circ$),
  which is consistent with the result given by \citet{tom96}.
However, Fig. \ref{fig-ang}(c, d) clearly shows an enhanced concentration of
  the galaxies with SF/AGN activity along the line between the two subclusters ($315^\circ-45^\circ$).
In A1750, there are few galaxies in the region between A1750 C and N,
  and there are few emission-line galaxies associated with subclusters.

%%%%%%%%%%%%%%%%%%%%%%%%%%%%%%%%%%%%
% Figure 12
%%%%%%%%%%%%%%%%%%%%%%%%%%%%%%%%%%%%
\begin{figure*}
\begin{center}
\includegraphics [width=160mm] {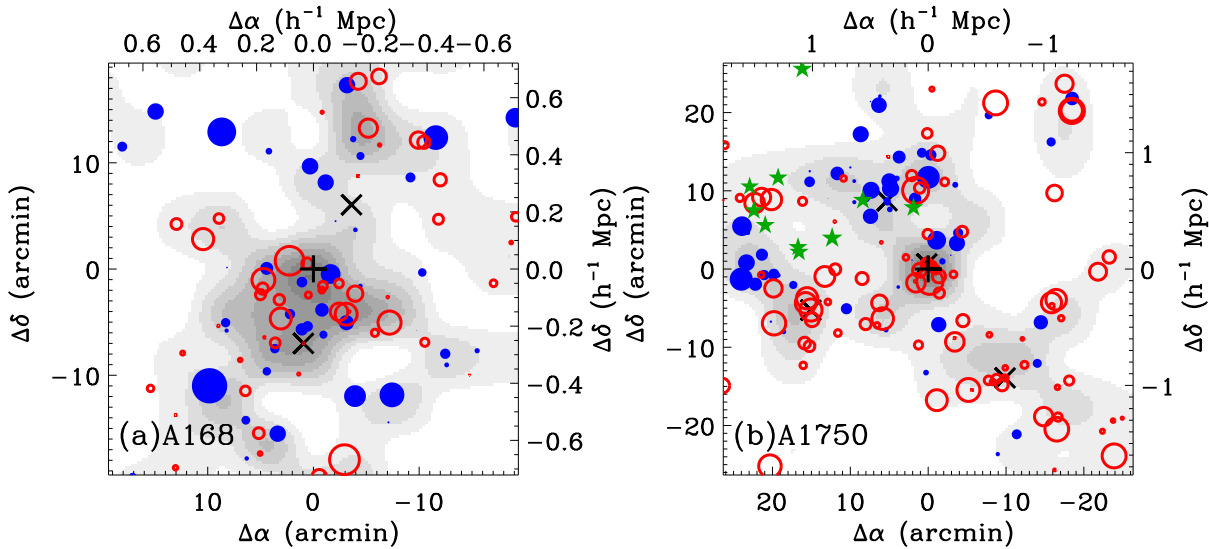}
\end{center}
\caption{Spatial distribution of the cluster galaxies overlaid on the galaxy number density maps
  for A168 ({\it a}) and A1750 ({\it b}). 
Open circles represent the cluster galaxies whose velocities are greater than the systemic velocity of their cluster, %  are plotted with open circles, 
while filled symbols represent those whose velocities are smaller  than the systemic velocity.
%  are shown by filled symbols.
The symbol size is proportional to the velocity deviation from the systemic velocity of the cluster.
In A1750 ({\it b}), star symbols represent 
the galaxies with radial velocities of $cz\sim22,500$ km s$^{-1}$
  that show a large velocity deviation ($cz-\overline{cz}\sim-3000$ km s$^{-1}$) from the main body.
%  are plotted by star symbols.
Plus sign indicates the centre we adopted, and crosses indicate the brightest galaxies in the subclusters.
North is up, and east is to the left.
}\label{fig-vel}
\end{figure*}

%%%%%%%%%%%%%%%%%%%%%%%%%%%%%%%%%%%%

\section{Discussion}\label{discuss}

\subsection{Effect of Cluster Merging on Galaxy Activity in A168}\label{dis-a168}

We found that the two subclusters in A168 
appear to have experienced last encounter about $3.5$ $h^{-1}$ Gyr ago,
  and to be now coming together from the maximum expansion of $1.58$ $h^{-1}$ Mpc about $0.4$ $h^{-1}$ Gyr ago.
It is needed to check the validity 
of the assumptions of two-body models (e.g., merging with no angular momentum).
%   may  not be justified in this study.
If the merging between two subclusters occurs with angular momentum (e.g., off-axis merger),
  the observed X-ray brightness will be distorted along the orbital motion
  as shown in the simulated one (e.g., Fig. 6 in \citealt{poo06}).
However, the observed {\it Chandra} X-ray image of A168
  is similar to the case with low angular momentum in the simulated one,
  which supports our assumption of radial orbits in two-body model for A168.

We found no enhanced concentration of blue galaxies 
  at the region between the two subclusters in A168, 
  which is consistent with the result in \citet{tom96}.
There are 55 galaxies in common between this study and \citet{tom96},
  if we use a matching tolerance of 15$\arcsec$.
Among 55 galaxies, 48 galaxies are found to be genuine cluster galaxies associated with A168
  when we use the velocity information provided by SDSS and NED.
Namely, the photometric sample of galaxies in \citet{tom96} includes 
  about 87 per cent of genuine cluster galaxies associated with A168.
In the sample of 48 genuine cluster galaxies associated with A168,
  we find that the colour types of 42 galaxies are assigned consistently,
  which leads to the similar finding in both studies
  [i.e., red (blue or semi-blue) galaxies determined in \citet{tom96} are 
  classified as red (blue) galaxies in this study].

When we use the emission-line diagnostics to detect galaxy activity,
  we find an enhanced concentration of the galaxies with SF/AGN activity at the region
  between the two subclusters in A168 (see Figs. \ref{fig-active}b and  \ref{fig-ang}d).
%The use of 
Emission-line diagnostics is more sensitive to probe SF/AGN activity
  compared to the integrated colour \citep{ken98,pc09} so that %Th  seems to help us to 
we could directly detect galaxy activity that was not found previously.
%There are two blue galaxies that do not show activity in the emission-line diagnostics,
%  and there are two galaxies having SF/AGN activity whose colour are not blue.
%  We also adopted the emission-line diagnostics
%   to probe the SF/AGN activity clearly \citep{pc09}.
%Therefore, the previous results that used the photometric sample of galaxies
%  appear to be affected by the non-member galaxies and 
%  by the use of total colour that is insensitive to SF/AGN activity compared to emission-line diagnostics.
Considering  the morphology-density relation 
%that  the late-type galaxies usually avoid high density regions,
the existence of the late-type galaxies with SF/AGN activity in the region between two subclusters of A168
%  where the galaxy number density is high,
  seems to be caused by some events such as the cluster merger or interaction.
The existence of the galaxies with SF/AGN activity in the region between two subclusters
  could be also a result of overlapping of the outskirts of two subclusters located 
  at slightly different distance along the line of sight.
If so, it is expected that there will be 
no angular variation of the fraction of the galaxies with SF/AGN activity
  as shown in Fig. \ref{fig-ang}. %, is expected.
However, Fig. \ref{fig-ang}(c, d) clearly shows an enhanced concentration of
  the galaxies with SF/AGN activity along the line between the two subclusters ($315^\circ-45^\circ$),
  implying that this scenario is less probable.

\subsection{Merging History of A1750}\label{dis-a1750}

Two major subclusters (A1750 C and N) appear to have started interaction and 
  to be coming together for the first time.
In A1750, there are few galaxies in the region between A1750 C and N,
  and there are few emission-line galaxies associated with subclusters.
This indicates that the current merger event started recently,
  not yet triggering any activity in cluster galaxies.

Detailed {\it XMM-Newton} images of A1750 C imply that it underwent a merger or interaction
 in the past $1-2$ Gyrs, and began to be in re-equilibrium \citep{bel04}.
Since the gas distribution to the southwestern side of A1750 C appears
  to be elongated in the direction of A1750 S, 
and  A1750 S is located at the line connecting A1750 C and N,
  A1750 S can be suspected to be a companion cluster responsible for the past interaction of A1750 C.
A1750 S lies at a projected distance of $\sim1.2$ $h^{-1}$ Mpc from
  A1750 C, and has a mean radial velocity similar to that of A1750 C.
The fact that we cannot find acceptable solutions in the bound region
  for the two-body model of A1750 N and C,
  when we use the relative radial velocity  between two brightest galaxies in subclusters,
  may support that A1750 C experienced a merger or interaction in the past.
It is noted that the existence of no acceptable solutions in the bound region
  may be also due to the wrong assumption of two-body model (e.g., merging with no angular momentum).
However, it is difficult to determine whether the interaction between A1750 N and C
  is similar to an off-axis merging by comparing with simulated X-ray images,
  because the distortion of X-ray contours is hardly seen.

On the other hand, the past interaction of A1750 C may be related to the other subcluster.
There is seen a concentration of galaxies to the east of A1750 C
  in the galaxy number density map (shown by a dashed ellipse and denoted by `E' in Fig. \ref{fig-sub}d),
  though X-ray images do not show any noticeable peaks.
The velocity distribution of galaxies in A1750 E is far from Gaussian,
  and the velocity dispersion and virial mass for A1750 E are abnormally large,
indicating  that it may not be a genuine subcluster currently.
The past interaction of A1750 E with A1750 C is suspected by the facts that 
  the compression of X-ray image is to the direction of east \citep{bel04}, and
  that A1750 E is much more smoothly connected to A1750 C compared to A1750 S
  as shown in the galaxy number density map.
Interestingly,
  there is one active galaxy on the line connecting A1750 E and C,
  and there are two active galaxies (one early type and one late type)
  just to the northeast in A1750 E.
It is difficult to determine whether the activity of
  the galaxies associated with A1750 E is triggered by a past interaction with A1750 C or not
  due to the small number of active galaxies.

A1750 E may be also related to A1750 N.
The galaxy number density map in Fig. \ref{fig-sub}(d)
  shows that A1750 C is largely connected to A1750 N through A1750 E.
To account for the velocity information in addition to the spatial distribution,
  we present the spatial distribution of cluster galaxies with measured velocities
  in Fig. \ref{fig-vel}.
The radial velocity of southern part of A1750 E is similar to that of A1750 C,
  and that of northern part of A1750 E is similar to that of A1750 N.

A1750 N is suspected to experience no mergers in the past, but to be weakly
  interacting with other sources other than A1750 C now. 
This is consistent with the findings based on  XMM-Newton data by \citet{bel04}:
 (1) the X-ray image of A1750 N is elongated along the northeast direction;
 (2) there is another extended X-ray source just to the north of A1750 N;
 (3) the gas temperature of A1750 N is uniform; and
 (4) the cooling flow is not disrupted in the core region of A1750 N.
The distribution of galaxies in the number density and 
  spatial-velocity maps (see Fig. \ref{fig-vel}) indicate that
the current interaction/accretion of A1750 N may be related to A1750 E.
In addition, if we plot the spatial distribution of
  the galaxies with $cz\sim$22,500 km s$^{-1}$ that might be another small group 
  (star symbols in Fig. \ref{fig-vel}),
  the spatial distribution of these galaxies is overlapped 
  with the region connecting to A1750 N and A1750 E.
In conclusion, A1750 E maybe a subcluster that was partially disrupted 
  during the previous interaction with A1750 C 
  and is currently interacting with A1750 N.

%It is noted that the velocity dispersions and the virial masses
%  for the subclusters can be strongly affected by the interlopers
%  due to the small number statistics.
%For example, the velocity dispersion and the virial mass for A1750 E
%  is the largest among A1750 subclusters, though their velocity distribution
%  of galaxies is far from Gaussian.
%When we increase twice the radius of the circle that defines the subcluster of A1750 E,
%  the velocity distribution is still not Gaussian.
%It indicates that A1750 E is not a genuine subcluster,
%  therefore, the velocity dispersion and the virial mass is not meaningful.

\subsection{E+A Galaxies}\label{dis-e+a}

We could not find any `E+A' galaxies that have strong Balmer absorption lines
  with no [O II] emission line in either of A168 and A1750.
`E+A'  galaxies are usually regarded as post-starburst galaxies that 
  have experienced starbursts within the last Gyr, and the activity has been abruptly truncated.
Since they were first discovered in distant clusters \citep{dg83}
  and were found more in clusters than in the field \citep{dre99,pog99,tran03,tran04},
  they were supposed to be cluster-related phenomena.
However, the local ($z\leq0.1$) `E+A' galaxies are rarely found in clusters \citep{dre87},
  and are usually found in the low density environment such as poor groups and the field 
  \citep{zab96,qui04,bla04,goto05,yan08}.
Therefore, their origin and relation with clusters are still debated \citep{yang08}.
Interestingly, some `E+A' galaxies are found in nearby merging clusters
 (Coma at $z\sim0.023$; \citealt{pog04}, and A3921 at $z\sim0.095$; \citealt{fer05}),
  but they do not appear to be related with the current merging event. 
 
However, in some cases, `E+A' galaxies are expected to exist in merging clusters
  if the increased ram-pressure of ICM can trigger SF activity 
  of cluster galaxies and subsequently quench it.
We checked the equivalent width (EW) of H$\delta$ absorption lines
  for all galaxies in Fig. \ref{fig-active}.
It is found that the maximum values of EW(H$\delta$) are just 1.15 \AA~in A168 
and 1.17 \AA~in A1750
  that are much smaller than the typical criterion (5\AA) for `E+A' galaxies.
In conclusion, 
  the existence of galaxies with SF/AGN activity and the lack of `E+A' galaxies in A168,
  may indicate that the galaxy activity is induced by the cluster merger recently,
  and is not yet quenched.
For A1750, the activity of galaxies associated with the merging between two major subclusters
  might be not yet triggered, since the merging is in the early stage.

\section{Summary}\label{sum}

We present the results of a study of dynamical state for two merging binary clusters (A168 and A1750)
  and the activity of cluster galaxies using the SDSS galaxy sample of which redshifts
  are available in SDSS or NED.
Our primary results are summarised below.

\begin{enumerate}
\item We have found the substructures in each cluster,
  and have investigated the kinematic and photometric properties of the galaxies in the subclusters.

\item Using the two-body model analysis and the X-ray data,
  we have investigated the merger histories of A168 and A1750.
Two subclusters in A168 appear to have experienced last encounter about $3.5$ $h^{-1}$ Gyr ago,
  and to be now coming together from the maximum expansion of $1.58$ $h^{-1}$ Mpc about $0.4$ $h^{-1}$ Gyr ago.
Two major subclusters (A1750 C and N) appear to have started interaction and 
  to be coming together for the first time. 

\item We have found an excess of galaxies with SF/AGN activity in the region
  between the two subclusters of A168, which might have been triggered by the cluster merger.
However, we found no enhanced concentration of blue galaxies 
  at that region, which is consistent with the result in \citet{tom96}.

\item In A1750, we could not find any galaxies 
  that show strong activity in the region between 
  two subclusters (A1750 N and C), which is consistent with
  the scenario that they are in the early stage of merging.
A1750 E may be a subcluster that was partially disrupted 
  during the previous interaction with A1750 C 
  and is currently interacting with A1750 N.

\item We found no `E+A' galaxies in either of A168 or A1750.
\end{enumerate}
    
In conclusion, the cluster merger appears to trigger the activity of cluster galaxies, 
  and its effect is different depending on the merging stage.
Cluster mergers are usually in different stages, and
  the conditions of ICM and galaxies in these clusters are diverse. %not same.
Therefore it is needed to study more clusters
  in different stages of merging to understand  the relation between
  the galaxy activity and the dynamical state of the clusters.

\section*{Acknowledgments}
The authors thank the anonymous referee for useful comments that improved
  significantly the original manuscript.
This work was supported in part by a grant R01-2007-000-20336-0 
  from the Basic Research Program of the Korea Science and Engineering Foundation.

Funding for the SDSS and SDSS-II has been provided by the Alfred P. Sloan 
Foundation, the Participating Institutions, the National Science 
Foundation, the U.S. Department of Energy, the National Aeronautics and 
Space Administration, the Japanese Monbukagakusho, the Max Planck 
Society, and the Higher Education Funding Council for England. 
The SDSS Web Site is http://www.sdss.org/.

The SDSS is managed by the Astrophysical Research Consortium for the 
Participating Institutions. The Participating Institutions are the 
American Museum of Natural History, Astrophysical Institute Potsdam, 
University of Basel, Cambridge University, Case Western Reserve University, 
University of Chicago, Drexel University, Fermilab, the Institute for 
Advanced Study, the Japan Participation Group, Johns Hopkins University, 
the Joint Institute for Nuclear Astrophysics, the Kavli Institute for 
Particle Astrophysics and Cosmology, the Korean Scientist Group, the 
Chinese Academy of Sciences (LAMOST), Los Alamos National Laboratory, 
the Max-Planck-Institute for Astronomy (MPIA), the Max-Planck-Institute 
for Astrophysics (MPA), New Mexico State University, Ohio State University, 
University of Pittsburgh, University of Portsmouth, Princeton University,
the United States Naval Observatory, and the University of Washington. 

This research has made use of the NASA/IPAC Extragalactic Database (NED) 
which is operated by the Jet Propulsion Laboratory, California Institute of Technology, 
under contract with the National Aeronautics and Space Administration.

\label{lastpage}

\end{document}